\documentclass[preprintnumbers,amsmath,amssymb]{revtex4}

\usepackage{amsmath}  
\usepackage{amsfonts} 
\usepackage{graphicx} 

\begin{document}


\title{Theoretical and conceptual analysis of the celebrated $4\pi$-symmetry neutron interferometry experiments}
\author{Massimiliano Sassoli de Bianchi}
\affiliation{Laboratorio di Autoricerca di Base, 6914 Lugano, Switzerland}\date{\today}
\email{autoricerca@gmail.com} 

\begin{abstract}
\noindent In 1975, two experimental groups have independently observed the $4\pi$-symmetry of neutrons' spin, when passing through a static magnetic field, using a three-blade interferometer made from a single perfect Si-crystal (analogous to the Mach–Zehnder interferometer of light optics). In this article, we provide a complete analysis of the experiment, both from a theoretical and conceptual point of view. Firstly, we solve the Schr\"{o}dinger equation in the weak potential approximation, to obtain the amplitude of the refracted and forward refracted beams, produced by the passage of neutrons through one of the three plates of the LLL interferometer. Secondly, we analyze their passage through a static magnetic field region. This allows us to find explicit expressions for the intensities of the four beams exiting the interferometer, two of which will be interfering and show a typical $4\pi$-symmetry, when the strength of the magnetic field is varied. In the last part of the article, we provide a conceptual analysis of the experiment, showing a neutron's phase change, when passing through the magnetic field, is due to a longitudinal Stern-Gerlach effect, and not to a Larmor precession. We also emphasize that these experiments do not prove the observability of the sign change of the wave function, when a neutron is $2\pi$ rotated, but strongly indicate that the latter, like any other elementary ``particle,'' would be a genuinely non-spatial entity. 
\end{abstract}

\maketitle

\section{Introduction}
\label{intro}

In the eighties of last century, when I was still a student of physics at the University of Lausanne (Switzerland), one of my teachers, Prof. G. Wanders, proposed me to choose \emph{neutron interferometry} as one of the subjects of theoretical physics to prepare and present in the ambit of my study plan. This was the first time that I became aware of some of the most fascinating physics experiments ever made, truly fundamental for our understanding of quantum mechanics at a foundational level.

Following my degree, I went to the nearby University of Geneva, where for some time I was the assistant of Prof. C. Piron, well known for his contributions to quantum logic. I soon discovered that Piron was also fascinated by neutron interferometry experiments, as he considered they provided  evidence that microscopic quantum entities, like neutrons, were truly \emph{non-local}, as it was possible to create interference effects by acting only on one of the beams inside the interferometer, even when a single neutron at a time was present in it. 

Later on, I was back to Lausanne to start a PhD with Prof. Ph. A. Martin, at the Federal Institute of Technology. Among the themes of my research, there was the notion of time-delay in quantum scattering theory (one of the traditional subjects of research of the Swiss school of theoretical physics) and the related controversy over the so-called `tunneling times'. This led me to also investigate the spin rotation mechanism during the passage of a quantum particle through a magnetic field region (the so-called Larmor clock), with particular consideration of the role played by the field boundaries; and this brought me back, one more time, to the beautiful neutron interferometry experiments, and their demonstration that the $4\pi$-periodicity of the spinor wave function was perfectly real, and not a mere mathematical artefact. 

In more recent times, I was interested in the work of Prof. Diederik Aerts (a former student of Piron) in the foundations of quantum mechanics, and discovered that neutron interferometry experiments have also been fundamental in bringing him to take very seriously what the quantum formalism is telling us about the nature of the microscopic entities. Different from Piron, Aerts pushed the analysis of these experiments to their extreme logical consequences, concluding that non-locality is in fact to be interpreted as \emph{non-spatiality}, i.e., as the result of the fact that microscopic quantum entities, like neutrons, would generally `not be in space', but would be `brought into space' only when interacting with a measuring apparatus (or when forming macroscopic aggregates, in standard conditions). 

The present article was written with the idea of bringing together these different encounters of mine with neutron interferometry, to offer the reader a description of the passage of neutrons through a so-called LLL silicon crystal, when one of the refracted beams is subjected to the action of a local static magnetic field. This not only because neutron interferometry is a wonderful didactical tool to highlight interesting physics and some of the remarkable properties of quantum entities, but also because it can stimulate a deep reflection about their nature and the reasons of their strange behavior. This is why, in addition to the theoretical calculations, I will also provide a conceptual analysis of the experiment, particularly for what concerns the notion of non-locality/non-spatiality and the idea that by rotating an spin-${1\over 2}$ entity by $2\pi$, it would not come back to the same condition. 

More precisely, the article is organized as follows. In Sec.~\ref{angstrom}, I provide some basic information about neutron interferometry and the specific geometry of a LLL-crystal. In Sec.~\ref{infinite}, I solve the stationary Schr\"{o}dinger equation for a neutron in an infinite crystal, in the weak potential approximation, assuming that there is only one reciprocal lattice vector close to Ewald's sphere. In Sec.~\ref{half-infinite}, I use the infinite crystal solution to consider the situation of a neutron incident on a half-space crystal, and in Sec.~\ref{void}, I consider the passage of a neutron from the half-space crystal to the void. This allows me, in Sec.~\ref{finite}, to study the passage of a neutron through a finite crystal plate (one of the three plates, or blades, of the LLL interferometer). 

Our scope being that of studying the effect of a magnetic field, when locally applied on one of the two beams inside the interferometer, I consider, in Sec.~\ref{magnetic}, the problem of the passage of a neutron in a static homogeneous magnetic field, again remaining in the weak field approximation. Combining all these calculations, I am finally able, in Sec.~\ref{interference}, to provide a description of the four-beam output of the interferometer, and show that when the strength of the magnetic field is varied, the intensity of the two interfering beams exhibits a remarkable $4\pi$-modulation, confirmed by the historical experiments. 

The article then continues with a more conceptual analysis. In Sec.~\ref{larmor}, I ask if the dephasing introduced by the magnetic field can be interpreted as a Larmor precession, or should instead be attributed to a longitudinal Stern-Gerlach effect. In Sec.~\ref{rotation}, I also address the question if the sign change of a spin-${1\over 2}$ state, when rotated by $2\pi$, is truly observable, and if this is what the interferometry experiments with neutrons have really achieved. Then, in Sec.~\ref{non-spatiality}, I analyze the experiments from the viewpoint of non-locality, emphasizing that the latter should be interpreted as non-spatiality. Finally, in Sec.~\ref{conclusion} I offer a few conclusive remarks.

\section{Neutron interferometry}
\label{angstrom}

In 1967, Bernstein \cite{Bernstein1967}, and independently Aharonov and Susskind \cite{Aharonov1967}, proposed an experiment to observe the sign change of spin-${1\over 2}$ entities when $2\pi$-rotated. The theoretical basis for performing these experiments with neutron interferometers was given in 1976, by Eder and Zeilinger \cite{Eder1976}, exploiting the invention of perfect silicon crystal interferometers, by U. Bonse and M. Hart, in 1964, that was initially used only for X-rays \cite{Bonse1965}, but subsequently applied also to neutrons, by Rauch, Treimer and Bonse in 1974 \cite{Rauchetal1974}.

The celebrated experiments we are going to theoretically describe in this article are those conducted by Rauch \emph{et al} \cite{Rauchetal1975} and Werner \emph{et al} \cite{Werneretal1975} in 1975, where a LLL device made from a single Si-crystal (analogous to the Mach–Zehnder interferometer of light optics) was used to observe the $4\pi$-periodicity of the spinor wave function, when rotated by a static magnetic field. 

\begin{figure}[!ht]
\centering
\includegraphics[scale =0.19]{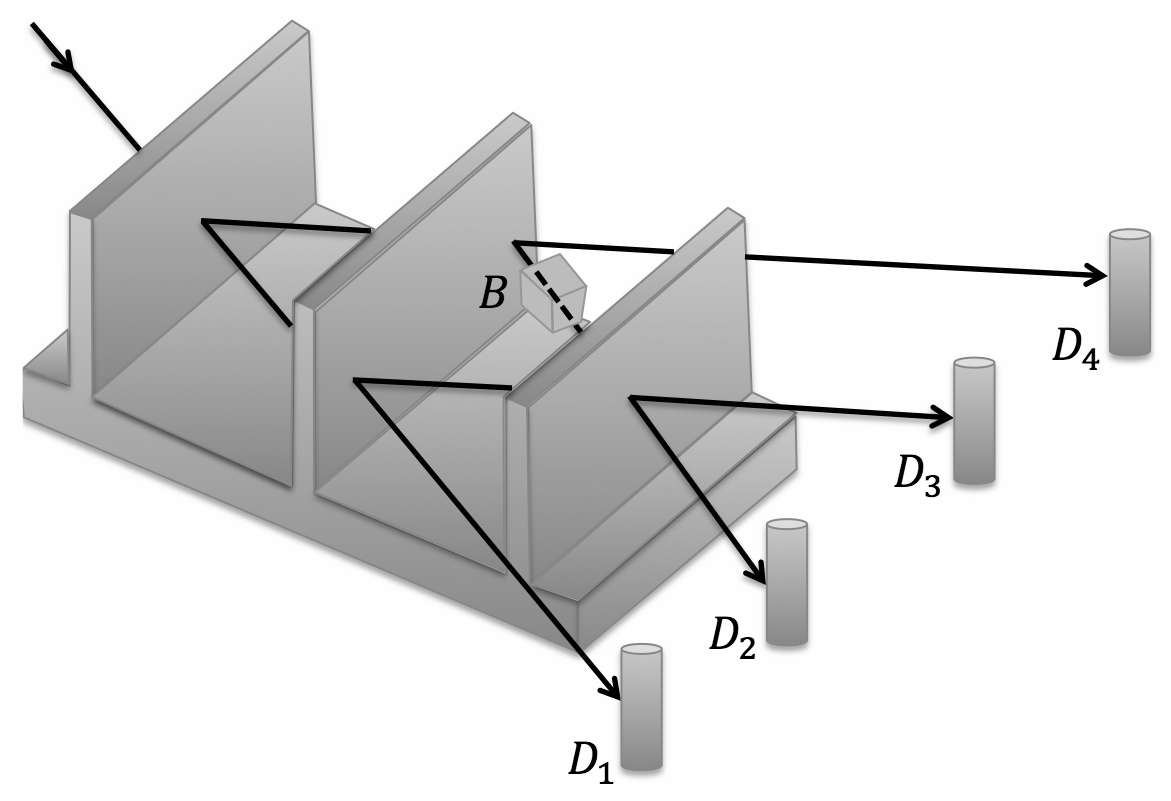}
\caption{A sketch of a monolithic perfect LLL crystal silicon (Si) interferometer, splitting the incident beam into four distinct beams, which are then detected by the detectors $D_1$, $D_2$, $D_3$ and $D_4$ (for instance, high pressure $^3{\rm He}$ neutron detectors). Along the path of one of the beams, a local and static magnetic field $B$ is applied, to induce a phase shift of part of the neutrons' wavefunction.  
\label{Monolitic}}
\end{figure}

As sketched in Fig.~\ref{Monolitic}, a LLL interferometer crystal is a monolithic device consisting of three perfect crystal plates (or blades) cut from a large and perfect ingot made of dislocation-free Si-crystal. The cut is done perpendicularly to a set of strongly reflecting Bragg planes [with Miller index (220)]. This corresponds to the so-called Laue (L) transmission geometry. The size of such LLL crystal is typically of 7 cm, and the thickness of the plates is less than half centimeter. 

When an incident, almost monochromatic beam of (thermal) ultracold neutrons (with energy $E\approx 0.025$~eV, and wavelength of about 2 \AA, comparable to the interatomic distance between atoms in solids), encounters the first plate, it splits into two distinct beams. Then, these two transmitted (forward refracted) and refracted (reflected by the Bragg's planes) beams also encounter the second plate and are further split into four distinct beams. Two of them exit the crystal, whereas the other two recombine (superpose) exactly at the level of the third plate, and finally also exit the interferometer. 

In other terms, the LLL device produces a split of the initial beam into four beams, two of them being non-interfering, and the other two being possibly interfering, if for instance some dephasing is produced by putting a `phase shifter' into one of the two beams, before their recombination in the third plate. In the present article, the phase shifter will be a static magnetic field, producing a rotation of the neutron's spin (see Fig.~\ref{Monolitic}). 

Neutron interferometry has been used in the last decades to test all kinds of basic features of quantum entities, thanks to the fact that neutrons are massive entities not too difficult to manipulate and detect, and that they can experience nuclear, electromagnetic as well as gravitational interactions. It is however not the scope of this article to review this vast and fascinating experimental literature, and we refer the interested readers to  \cite{BonseRauch1979,Hasegawa2011,Rauch2012,RauchWerner2015}, and the references cited therein. 

\section{Neutron in an infinite crystal}
\label{infinite}

To analyze the passage of a neutron through the silicon LLL crystal, we start by considering the stationary Schr\"{o}dinger equation in an infinite crystal: 
\begin{equation}
\left[-{\hbar^2\over 2m}\Delta + V({\bf x})\right]\psi({\bf x})=E\psi({\bf x}),
\label{Schroedinger-equation-crystal}
\end{equation}
where $E$  and $m$ are the energy and mass of the neutron, respectively, and $V({\bf x})=V({\bf x}+{\bf a}_i)$ is the periodic potential function describing the infinite crystal, with the ${\bf a}_i$, $i=1,2,3$, being the base vectors of the elementary unit of the Bravais lattice. 

To solve (\ref{Schroedinger-equation-crystal}), one has to use Bloch's theorem, saying that solutions $\psi({\bf x})$ are the product of a plane wave and a function having the periodicity of the Bravais lattice (see for instance \cite{Ashcroft1976}):
\begin{equation}
\psi({\bf x})=u({\bf x})e^{i{\bf k}\cdot{\bf x}}, \quad u({\bf x})=u({\bf x}+{\bf r}),
\label{factorization}
\end{equation}
where ${\bf r} = n_1{\bf a}_1+n_2{\bf a}_2+n_3{\bf a}_3$, and the $n_i$, $i=1,2,3$, can take integer (positive and negative) values.

Introducing the Fourier representation of the periodic functions $u({\bf x})$ and $V({\bf x})$, we can write: 
\begin{equation}
u({\bf x})=\sum_{{\bf g}}\tilde{u}({\bf g})e^{i{\bf g}\cdot{\bf x}},\quad V({\bf x})=\sum_{{\bf g}}\tilde{V}({\bf g})e^{i{\bf g}\cdot{\bf x}},
\label{Fourier-representation}
\end{equation}
where the vectors ${\bf g}$ belong to the reciprocal lattice of the Bravais lattice of vectors ${\bf r}$ (the set of wave vectors ${\bf g}$ for which the plane waves $e^{i{\bf g}\cdot{\bf x}}$ have the periodicity of the Bravais lattice, i.e., satisfying $e^{i{\bf g}\cdot{\bf r}}=1$), which can be written as: ${\bf g} = n_1{\bf g}_1+n_2{\bf g}_2+n_3{\bf g}_3$, with ${\bf g}_i = 2\pi V_e^{-1}\epsilon_{ijk}\, {\bf a}_i\wedge {\bf a}_j$, $i=1,2,3$, with $V_e={\bf a}_1\cdot ({\bf a}_2 \wedge{\bf a}_3)$ being the volume of the elementary unit. 

Inserting (\ref{Fourier-representation}) into (\ref{Schroedinger-equation-crystal}), one obtains the momentum space representation of the Schr\"{o}dinger equation: 
\begin{equation}
\left[{\hbar^2\over 2m}\| {\bf k}+{\bf g}\|^2 -E\right]\tilde{u}({\bf g})=-\sum_{{\bf g}'}\tilde{V}({\bf g}-{\bf g}')\tilde{u}({\bf g}').
\label{Schroedinger-equation-crystal2}
\end{equation}
To describe the interaction of the neutron with the nuclei (neutron's interaction with matter being dominated by the strong neutron-nuclear interaction) it is sufficient to consider the Fermi pseudopotential \cite{Fermi1936}:
\begin{equation}
V({\bf x})=\lambda \sum_{n} \delta\left({\bf x}-{\bf r}_n\right),
\label{pseudopotential}
\end{equation}
where ${\bf r}_n$ is the position of the $n$-th nucleus, and $\lambda = {2\pi \hbar\over m}b_c$, where the parameter $b_c$, characterizing the neutron-nuclear interaction, is called the scattering length (which is typically in the range of $-5$ to $10$ fermi, but in most cases is positive, as it will assumed here). Clearly, a potential of this form describes a situation where the interactions between the neutron and the nuclei are conveyed by forces which are of almost zero range with respect to the angstr\"{o}m scale, which is that of the phenomena we are here interested. Indeed, the range of the strong nuclear force (which is typically that of the nuclear radius) is much smaller than the de Broglie wavelength of the ultracold neutrons. 

To study (\ref{Schroedinger-equation-crystal2}), we will assume that the potential's Fourier transform $\tilde{V}({\bf g})$ is small, so that in a first approximation we can write: 
\begin{equation}
\left[{\hbar^2\over 2m}\| {\bf k}+{\bf g}\|^2 -E\right]\tilde{u}({\bf g})\approx 0.
\label{Schroedinger-equation-crystal3}
\end{equation}
This means that $\tilde{u}({\bf g})$ can be sensibly different from zero only for those values of ${\bf g}$ such that:
\begin{equation}
{\hbar^2\over 2m}\| {\bf k}+{\bf g}\|^2 \approx E,
\label{Condition}
\end{equation}
and of course this statement is meaningful only if such condition is verified for at least one ${\bf g}$. On that respect, we observe that the factorization (\ref{factorization}) is not unique, as one can also write: 
\begin{equation}
\psi({\bf x})=u'({\bf x})e^{i{\bf k}'\cdot{\bf x}}, \,\, u'({\bf x})=u({\bf x})e^{-i{\bf g}_0\cdot{\bf x}},\,\, {\bf k}'={\bf k}+{\bf g}_0.
\label{factorization2}
\end{equation}
Therefore, we can always choose ${\bf k}$ so that it obeys (\ref{Condition}), for instance for the value ${\bf g}={\bf 0}$, and for such choice $\tilde{u}({\bf 0})$ will be sensibly different from zero. 

\begin{figure}[!ht]
\centering
\includegraphics[scale =0.2]{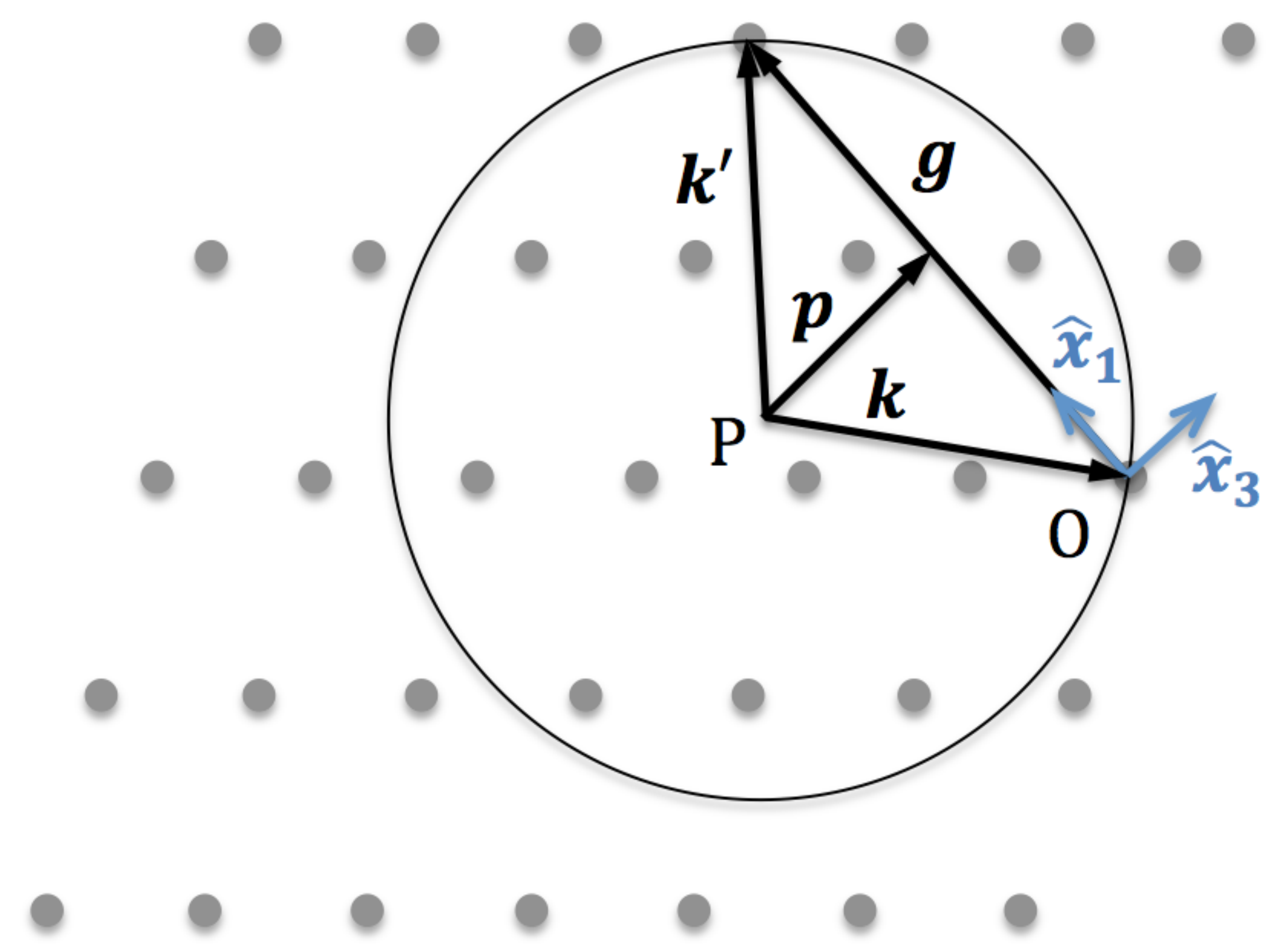}
\caption{Ewald's sphere construction, with $P$ the center of the sphere, $O$ the origin of the reciprocal space, $\vec{PO}= {\bf k}$, and ${\hbar^2\over 2m}\| {\bf k}\|^2 = {\hbar^2\over 2m}\| {\bf k}'\|^2= {\hbar^2\over 2m}\| {\bf k}+{\bf g} \|^2 =E$.
\label{Ewald}}
\end{figure}

Then, can we find a ${\bf g}\neq{\bf 0}$ such that ${\hbar^2\over 2m}\| {\bf k}+{\bf g}\|^2\neq 0$? To this end, we can use Ewald's construction in the reciprocal lattice, which consists in drawing a sphere of radius $\| {\bf k}\|$, as in Fig.~\ref{Ewald}. If, apart from the origin, no reciprocal lattice point lies on the surface of the sphere, then $\tilde{u}({\bf 0})$ is the only non negligible amplitude and we will only have a one-wave mode: 
\begin{equation}
\psi({\bf x})\approx \tilde{u}({\bf 0}) e^{i{\bf k}\cdot{\bf x}}.
\label{onewavemode}
\end{equation}
On the other hand, assuming that there is a unique vector ${\bf g}$ describing a lattice point that lies on (or is very close to) the sphere, then there is a vector ${\bf k}'$ satisfying Laue's diffraction condition ${\bf k}'-{\bf k} ={\bf g}$ (which says that constructive interference can occur provided that the change in wave vector is a vector of the reciprocal lattice), and we are in the situation of a two-wave mode: 
\begin{equation}
\psi({\bf x})\approx \tilde{u}({\bf 0}) e^{i{\bf k}\cdot{\bf x}} + \tilde{u}({\bf g}) e^{i({\bf k}+{\bf g})\cdot{\bf x}}.
\label{twowavemode}
\end{equation}
Eq.~(\ref{Schroedinger-equation-crystal2}) then reduces to the system of two equations: 
\begin{eqnarray}
&\left[{\hbar^2\over 2m}\| {\bf k}\|^2 -E + \tilde{V}({\bf 0})\right]\tilde{u}({\bf 0}) + \tilde{V}(-{\bf g})\tilde{u}({\bf g})=0,\nonumber\\
&\tilde{V}({\bf g})\tilde{u}({\bf 0}) + \left[{\hbar^2\over 2m}\| {\bf k}+{\bf g} \|^2 -E + \tilde{V}({\bf 0})\right]\tilde{u}({\bf g}) =0.
\label{Schroedinger-equation-crystal3}
\end{eqnarray}

Writing the above in matrix form: $A\tilde{{\bf u}}=0$, with $\tilde{{\bf u}}=(\tilde{u}({\bf 0}),\tilde{u}({\bf g}))$, and being the equation homogeneous, it is clear that a non-trivial solution can be obtained only if the determinant of the $2\times 2$ matrix $A$ is zero. This gives the dispersion relation: 
\begin{equation}
\left[{\hbar^2\over 2m}\| {\bf k}\|^2 -E + \tilde{V}({\bf 0})\right]\left[{\hbar^2\over 2m}\| {\bf k}+{\bf g} \|^2 -E + \tilde{V}({\bf 0})\right] = |\tilde{V}({\bf g})|^2,
\label{detzero}
\end{equation}
where we have used that $\tilde{V}(-{\bf g})=\tilde{V}^*({\bf g})$, considering that the potential is assumed to be real (no absorption). 

At this point, we look for conditions such that the two waves of the two-wave mode (\ref{twowavemode}) have amplitudes of the same order of magnitude, i.e., $\tilde{u}({\bf 0})/\tilde{u}({\bf g})=O(1)$. According to (\ref{Schroedinger-equation-crystal3}), this is the case if: 
\begin{eqnarray}
&{\hbar^2\over 2m}\| {\bf k}\|^2 -E + \tilde{V}({\bf 0}) =O(\tilde{V}),\label{magn}\\
&{\hbar^2\over 2m}\| {\bf k}+{\bf g} \|^2 -E + \tilde{V}({\bf 0})=O(\tilde{V}).
\label{magnitude}
\end{eqnarray}
For a given $E$, the above conditions for ${\bf k}$ are not necessarily satisfied if ${\bf k}$ only satisfies (\ref{detzero}). By setting ${\bf k}= {\bf p} - {1\over 2} {\bf g}$, (\ref{magn})-(\ref{magnitude}) become: 
\begin{eqnarray}
&{\hbar^2\over 2m}\| {\bf p} - {1\over 2} {\bf g}\|^2 -E + \tilde{V}({\bf 0}) =O(\tilde{V}),\label{magnitude2a}\\
&{\hbar^2\over 2m}\| {\bf p} + {1\over 2} {\bf g} \|^2 -E + \tilde{V}({\bf 0})=O(\tilde{V}).
\label{magnitude2b}
\end{eqnarray}
Considering (\ref{magnitude2a})$+$(\ref{magnitude2b}) and (\ref{magnitude2a})$-$(\ref{magnitude2b}), we can also write: 
\begin{eqnarray}
&{\hbar^2\over 2m}\left(\| {\bf p}\|^2 + {1\over 4} \|{\bf g}\|^2\right) -E =O(\tilde{V}),\label{magnitude3a}\\
&{\hbar^2\over 2m} {\bf p}\cdot {\bf g}=O(\tilde{V}).
\label{magnitude3b}
\end{eqnarray}

We observe that (\ref{detzero}) implies (\ref{magnitude3a}) if (\ref{magnitude3b}) is satisfied, which means that we have to choose ${\bf p}$ to be orthogonal (or almost orthogonal) to ${\bf g}$. To see this, we consider a system of coordinates generated by three orthonormal vectors ${\hat{\bf x}}_1$, ${\hat{\bf x}}_2$ and ${\hat{\bf x}}_3$, such that ${\bf g}$ is parallel to ${\hat{\bf x}}_1$, i.e., ${\bf g}=g_1 {\hat{\bf x}}_1$, and such that ${\bf k}$ and ${\bf p}$ belong to the same plane generated by ${\hat{\bf x}}_1$ and ${\hat{\bf x}}_3$, with $k_3 = {\bf k}\cdot {\hat{\bf x}}_3 = {\bf p}\cdot{\hat{\bf x}}_3 =p_3 >0$ (see Fig.~\ref{Ewald}). We have: $k_1 = {\bf k}\cdot {\hat{\bf x}}_1=({\bf p} - {1\over 2} {\bf g}) \cdot {\hat{\bf x}}_1=p_1 - {1\over 2}g_1$, so that $\| {\bf k}\|^2= k_1^2+k_3^2= (p_1 - {1\over 2}g_1)^2 + k_3^2= p_1^2+k_3^2 -p_1g_1 + {1\over 4}g_1^2 = \| {\bf p}\|^2+ {1\over 4} \|{\bf g}\|^2 -p_1g_1$. Similarly, $\| {\bf k}+{\bf g} \|^2 =(p_1 + {1\over 2}g_1)^2 + k_3^2=\| {\bf p}\|^2+ {1\over 4} \|{\bf g}\|^2 +p_1g_1$. Inserting these expressions in (\ref{detzero}), setting $k_0^2\equiv {2mE\over \hbar^2}$ and $\tilde{U}\equiv {2m\tilde{V}\over \hbar^2}$, then using the identity $(a-b)(a+b)=a^2-b^2$, we find: 
\begin{equation}
\left[\| {\bf p}\|^2+ {1\over 4} \|{\bf g}\|^2 -k_0^2+\tilde{U}({\bf 0})\right]^2 = (p_1g_1)^2 + |\tilde{U}({\bf g})|^2,
\label{detzero-bis}
\end{equation}
which clearly implies (\ref{magnitude3a}) if (\ref{magnitude3b}) is satisfied, i.e., if $p_1g_1=O(\tilde{U})$.

We see from (\ref{detzero-bis}) that the compatibility equation (\ref{detzero}) has two solutions such that $k_3=p_3>0$, which are the following: 
\begin{equation}
k_3^2=k_0^2-{1\over 4}g_1^2-p_1^2-\tilde{U}({\bf 0})\mp \sqrt{(p_1g_1)^2 + |\tilde{U}({\bf g})|^2}.
\label{detzero-solutions}
\end{equation}
In other terms, for a given $k_1$, there are two two-wave modes, which we will denote ${\bf k}_1$ and ${\bf k}_2$ (something that could have been deduced by observing that the dispersion relation (\ref{detzero}) is a quadratic equation). Because of (\ref{magnitude3b}), we can neglect the $p_1^2$ term in (\ref{detzero-solutions}) and write: 
\begin{equation}
k_{{1\atop 2},3}={\bf k}_{1\atop 2}\cdot \hat{\bf x}_3= \sqrt{k_0^2-{1\over 4}g_1^2}-{1\over 2}{\tilde{U}({\bf 0})\over \sqrt{k_0^2-{1\over 4}g_1^2}} \mp{1\over 2}\sqrt{(p_{{1\atop 2},1}\, g_1)^2 + |\tilde{U}({\bf g})|^2\over k_0^2-{1\over 4}g_1^2}+O(\tilde{U}^2).
\label{detzero-solutions2}
\end{equation}

It is useful to define a dimensionless parameter $y$ such that: 
\begin{equation}
{\bf p}\cdot {\bf g}=p_1g_1=y\, |\tilde{U}({\bf g})|.
\label{y}
\end{equation}
This means that $y$ measures the deviation from orthogonality between ${\bf p}$ and ${\bf g}$, thus fixing the value of $k_1$. Eq~(\ref{detzero-solutions}) then becomes:
\begin{equation}
k_{{1\atop 2},3}= \sqrt{k_0^2-{1\over 4}g_1^2}\left[1 - {1\over 2}{1\over k_0^2-{1\over 4}g_1^2}\left(\tilde{U}({\bf 0})\pm \sqrt{1+y^2}\, |\tilde{U}({\bf g})|\right)\right]+ O(\tilde{U}^2).
\label{detzero-solutions2}
\end{equation}

To determine the amplitudes of these two modes, we have to solve (\ref{Schroedinger-equation-crystal3}). For this, we need to calculate $\|{\bf k}_1\|^2$ and $\|{\bf k}_2\|^2$. According to (\ref{detzero-solutions2}), we have ($\sigma = 1,2$): 
\begin{eqnarray}
\|{\bf k}_{\sigma}\|^2 &=& k_{\sigma,3}^2 - p_{\sigma,1}g_1 + {1\over 4}g_1^2 + p_{\sigma,1}^2\nonumber\\
&=&k_{\sigma,3}^2 - y\, |\tilde{U}({\bf g})| + {1\over 4}g_1^2 +O(\tilde{U}^2)\\
&=&k_0^2-\tilde{U}({\bf 0}) -|\tilde{U}({\bf g})|\left(y\pm \sqrt{1+y^2}\right)+O(\tilde{U}^2).\nonumber
\label{k-square}
\end{eqnarray}
Using (\ref{k-square}) into (\ref{Schroedinger-equation-crystal3}), we obtain: 
\begin{eqnarray}
\tilde{u}({\bf g}) &=& -\tilde{U}^*({\bf g})^{-1}\left(\|{\bf k}\|^2 -k_0^2 + \tilde{U}({\bf 0}) \right)\tilde{u}({\bf 0})\nonumber\\
&=&e^{i\phi}\left(y\pm \sqrt{1+y^2}\right)+O(\tilde{U}),
\label{ug}
\end{eqnarray}
where we have set: $e^{i\phi}\equiv |\tilde{U}({\bf g})|^{-1}\tilde{U}({\bf g})$. We can thus write for the two two-wave modes:
\begin{eqnarray}
&&\tilde{u}_\sigma({\bf g})= X_\sigma\tilde{u}_\sigma({\bf 0})+O(\tilde{U}),\quad \sigma = 1,2,\nonumber\\
&&X_{1\atop 2}=e^{i\phi}\left(y\pm \sqrt{1+y^2}\right).
\label{ug2}
\end{eqnarray}

In other terms, for a given $k_1$ (i.e., a given $y$), in a first approximation we have in the crystal the two two-wave modes ($\sigma = 1,2$):
\begin{eqnarray}
\psi_\sigma({\bf x})&\approx& \tilde{u}_\sigma({\bf 0}) e^{i{\bf k}_\sigma\cdot{\bf x}} + \tilde{u}_\sigma({\bf g}) e^{i({\bf k}_\sigma+{\bf g})\cdot{\bf x}}\nonumber\\
&\approx&\tilde{u}_\sigma({\bf 0})(1+X_\sigma e^{i{\bf g}\cdot{\bf x}})e^{i{\bf k}_\sigma\cdot{\bf x}},
\label{twowavemode2}
\end{eqnarray}
which means that the neutron's wave function inside the crystal consists in a coherent superposition of four plane waves of wave vectors ${\bf k_1}$, ${\bf k_2}$, ${\bf k_1}+ {\bf g}$ and ${\bf k_2}+ {\bf g}$, with the two wave vectors ${\bf k_1}$, ${\bf k_2}$ that are nearly collinear (and give rise to interference ``beats,'' called Pendell\"{o}sung interference fringes, on a macroscopic scale of approximately $100$ $\mu{\rm m}$).

We observe that if the potential is symmetric, i.e., $V({\bf x})=V(-{\bf x})$, then $\tilde{V}({\bf g})$ is a real function and $e^{i\phi}=1$. Also, in the situation depicted in Fig.~\ref{Ewald}, where $y=0$, we have $X_{1\atop 2}=\pm 1$, so that:
\begin{equation}
\psi_\sigma({\bf x})\approx \tilde{u}_\sigma({\bf 0}) (1\pm e^{i{\bf g}\cdot{\bf x}})e^{i{\bf k}_\sigma\cdot{\bf x}},\quad \sigma = 1,2.
\label{twowavemode3}
\end{equation}
This means that $|\psi_1({\bf x})|$ is maximal (respectively, $|\psi_2({\bf x})|$ is minimal) at the vertices ${\bf x} = n_1{\bf a}_1+n_2{\bf a}_2+n_3{\bf a}_3$ of the periodic structure occupied by the nuclei. Therefore, if there would be absorption, it would only affect the first mode. However, since we know that thermal neutrons are practically not absorbed by a Si-crystal, both two-wave modes propagate within it (something that we have implicitly assumed by considering a real potential function).

\section{Neutron incident on a crystal occupying a half-space}
\label{half-infinite}

We consider now the situation of a neutron incident on an infinite crystal occupying the half space $\{{\bf x}\in{\mathbb R}^3 | x_3={\bf x}\cdot{\hat{\bf x}}_3 >0\}$. We look for conditions such that the incident neutron can produce within the crystal a two-wave mode. For this, we have to impose continuity of the wave function at the $x_3=0$ interface of the crystal, as well as of the wave function's derivative along the ${\hat{\bf x}}_3$-direction. 

\begin{figure}[!ht]
\centering
\includegraphics[scale =0.16]{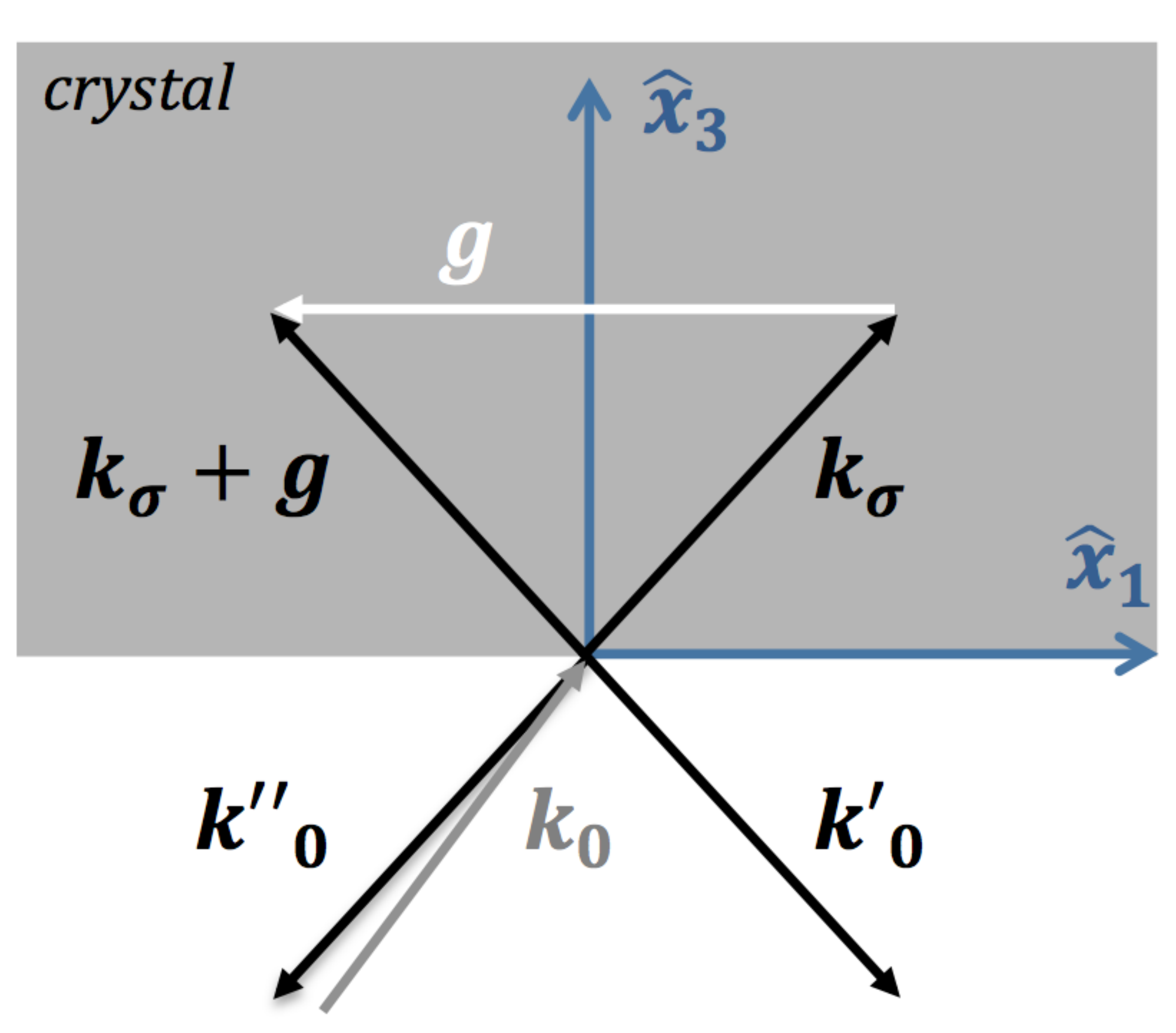}
\caption{The crystal occupying the $x_3>0$ half-space, with the incident, reflected and diffracted wave vectors. 
\label{half-space}}
\end{figure}

To do this, we assume that ${\bf g}$ is parallel to the crystal's interface, and more specifically that ${\bf g}=g_1 {\hat{\bf x}}_1$ (as we assumed in the previous section). Then, if ${\bf k}_0$ is the wave vector of the incident neutron, the connection conditions impose that $k_{0,1}=k_{1,1}=k_{2,1}$. In other terms, $k_1$ is fixed and we are in the same condition as in the previous analysis of the infinite crystal (where $k_1$ was determined by the parameter $y$). 
We also assume that the plan generated by ${\bf k}_0$ and ${\bf g}$ is perpendicular to the crystal's interface, so that we are in the situation depicted in Fig.~\ref{half-space}.

To satisfy the connection conditions, we introduce two reflected waves of wave vectors ${\bf k}'_0$ and ${\bf k}''_0$, with $\|{\bf k}_0\|=\|{\bf k}'_0\|=\|{\bf k}''_0\|$, and $k'_{0,1}=k_{0,1}$, $k''_{0,1}- k_{0,1}=g_1$. Denoting $\psi^-({\bf x})$ the solution for $x_3<0$, we thus write:
\begin{equation}
\psi^-({\bf x})= e^{i{\bf k}_0\cdot{\bf x}} + A\, e^{i{\bf k}'_0\cdot{\bf x}}+B\, e^{i{\bf k}''_0\cdot{\bf x}}.
\label{minussolution}
\end{equation}
For $x_3>0$, the solution is a superposition of the two two-wave modes $\psi_1({\bf x})$ and $\psi_2({\bf x})$, so that we can write: 
\begin{equation}
\psi^+({\bf x})\approx \sum_{\sigma =1,2} \tilde{u}_\sigma (1+X_\sigma e^{i{\bf g}\cdot{\bf x}})e^{i{\bf k}_\sigma\cdot{\bf x}},
\label{plussolution}
\end{equation}
where $\tilde{u}_\sigma \equiv \tilde{u}_\sigma({\bf 0})$. The continuity of the wave function at $x_3=0$ gives:
\begin{equation}
1+A+B\, e^{i g_1 x_1}=\tilde{u}_1 +\tilde{u}_2 + (\tilde{u}_1X_1+\tilde{u}_2X_2)\, e^{i g_1 x_1},
\label{continuity}
\end{equation}
which implies that: 
\begin{equation}
\tilde{u}_1 +\tilde{u}_2 = 1+A, \quad \tilde{u}_1X_1+\tilde{u}_2X_2 = B.
\label{continuity2}
\end{equation}
Similarly, the continuity of the $x_3$-partial derivative of the wave function at $x_3=0$ gives:
\begin{eqnarray}
&&k_{1,3}\tilde{u}_1 +k_{2,3}\tilde{u}_2 = k_{0,3}(1-A),\nonumber\\
&&k_{1,3}\tilde{u}_1X_1+k_{2,3}\tilde{u}_2X_2 = -k_{0,3}B.
\label{continuity3}
\end{eqnarray}
Combining (\ref{continuity2}) with (\ref{continuity3}), we can eliminate $A$ and $B$ and find the system of equations:
\begin{eqnarray}
&&(k_{1,3}+k_{0,3})\tilde{u}_1 +(k_{2,3}+k_{0,3})\tilde{u}_2 = 2k_{0,3},\nonumber\\
&&(k_{1,3}+k_{0,3})X_1\tilde{u}_1 +(k_{2,3}+k_{0,3})X_2\tilde{u}_2 = 0,
\label{continuity4}
\end{eqnarray}
which has the solutions:
\begin{equation}
\tilde{u}_1 ={2k_{0,3}\over k_{1,3}+k_{0,3}}{X_2\over X_2-X_1},\quad \tilde{u}_2 ={2k_{0,3}\over k_{2,3}+k_{0,3}}{X_1\over X_1-X_2}.
\label{solution-u}
\end{equation}

At this point, we estimate $k_{0,3}$. We have $k_0^2= k_{0,1}^2 + k_{0,3}^2$ and $k_{0,1}=k_{\sigma,1}=p_{\sigma,1}+{1\over 2}g_1$, so that:
\begin{eqnarray}
k_{0,1}^2&=& {1\over 4}g_1^2 + p_{\sigma,1}g_1+ p_{\sigma,1}^2\\
&=& {1\over 4}g_1^2 + y\, |\tilde{U}({\bf g})| + O(\tilde{U}^2),
\label{ko1}
\end{eqnarray}
where for the last equality we have used (\ref{y}). Thus, $k_{0,1}^2= {1\over 4}g_1^2 + O(\tilde{U})$, so that: 
\begin{equation}
k_{0,3}= \sqrt{k_0^2 -{1\over 4}g_1^2} + O(\tilde{U}).
\label{k03}
\end{equation}
Comparing (\ref{k03}) with (\ref{detzero-solutions2}), we see that the $k_{\sigma,3}$ are very close to $k_{0,3}$, in the sense that their difference is $O(\tilde{U})$. We can then rewrite (\ref{solution-u}) as follows: 
\begin{equation}
\tilde{u}_1 ={X_2\over X_2-X_1}[1+O(\tilde{U})],\quad \tilde{u}_2 ={X_1\over X_1-X_2}[1+O(\tilde{U})], 
\label{solution-u2}
\end{equation}
and using (\ref{ug2}), we can also write: 
\begin{equation}
\tilde{u}_{1\atop 2} ={1\over 2}\left(1\mp {y\over \sqrt{1+y^2}}\right)[1+O(\tilde{U})]. 
\label{solution-u3}
\end{equation}
It follows that $\tilde{u}_1+\tilde{u}_2=1+O(\tilde{U})$, so that, according to (\ref{continuity2}), $A= O(\tilde{U})$. We also find that $\tilde{u}_1X_1+\tilde{u}_2X_2 = O(\tilde{U})$, and from (\ref{continuity2}) we obtain $B= O(\tilde{U})$. In other terms, reflected waves are negligible at the first order in the potential. 

A remark is in order. It is possible to define a vector ${\bf k}_B$ such that $\|{\bf k}_B\|=\|{\bf k}_0\|$ and $k_{B,1}=-{1\over 2}g_1$. Clearly, such vector precisely obeys Laue's (or Bragg's) diffraction condition $\|{\bf k}_B\|=\|{\bf k}_B+{\bf g}\|$. Eq. (\ref{k03}) then tells us that ${\bf k}_0$ is very close to ${\bf k}_B$. So, to produce a two-wave mode inside the crystal, the incident wave vector must verify Laue's condition with very good approximation. 

Our previous definition (\ref{y}) of the parameter $y$ can be rewritten in terms of ${\bf k}_0$:
\begin{equation}
\left({\bf k}_0 + {1\over 2} {\bf g}\right)\cdot {\bf g}=\left({\bf k}_0 -{\bf k}_B\right)\cdot {\bf g}=y\, |\tilde{U}({\bf g})|.
\label{y-bis}
\end{equation}
In other terms, the characteristic parameter for dynamical diffraction $y$ measures the difference between ${\bf k}_0$ and ${\bf k}_B$, and in terms of the angles of incidence (measured for instance from the normal of the crystal's plane), we have that $y\propto (\theta_0 - \theta_B)$, i.e., $y$ is directly proportional to the deviation from the exact Bragg's angle condition (when such deviation is small).

\section{Passage from a half-space crystal to the void}
\label{void}

We now consider the situation where a two-wave mode, specified by $(E, {\bf k}_\sigma, {\bf g})$, with ${\bf k}_\sigma>0$, propagates inside a half-space crystal and reach the interface crystal-void. We assume that for $x_3<d$ we are in the crystal medium, whereas for $x_3>d$ we are in the void, with $d>0$. 

By similar arguments as those presented in the previous section, one finds that the incidence of the mode on the surface crystal-void will give rise to reflected waves, which are $O(\tilde{U})$, and therefore can be neglected in a first approximation. Thus, we can write: 
\begin{eqnarray}
\psi^-({\bf x})&\approx& \tilde{u}_\sigma (1+X_\sigma e^{i{\bf g}\cdot{\bf x}})e^{i{\bf k}_\sigma\cdot{\bf x}},\quad x_3<d \\
\psi^+({\bf x}) &=& C e^{i{\bf k}_0 \cdot{\bf x}} + D e^{i{\bf k}'_0 \cdot{\bf x}},\quad x_3>d,
\label{solutioncrystal-void}
\end{eqnarray}
where ${\bf k}_0$ describes the transmitted (forward-diffracted) wave, so that $k_{0,1}=k_{\sigma,1}$, and ${\bf k}'_0$ (not to be confused with the ${\bf k}'_0$ described in the previous section) describes the diffracted wave, so that $k'_{0,1}=k_{\sigma,1}+ g_1$.
So, the continuity condition at $x_1=d$ will fix $C$ and $D$, so that at the first order in $\tilde{U}$ we can write: 
\begin{equation}
\psi^+({\bf x})\approx \tilde{u}_\sigma \left[e^{i(k_{\sigma,3}- k_{0,3})d} e^{i{\bf k}_0 \cdot{\bf x}}+ 
e^{i(k_{\sigma,3}- k'_{0,3})d} e^{i{\bf k}'_0 \cdot{\bf x}}\right]
\end{equation}

\section{Passage through a finite plate}
\label{finite}

We then consider the situation of a crystal plate of thickness $d$, i.e., for $0<x_3<d$ we are in the crystal medium, whereas for $x_3<0$ and $x_3>d$ we are in the void. 

We know that in the crystal we have the superposition of two two-wave modes, produced by the incident neutron of wave vector ${\bf k}_0$, as described in Sec.~\ref{infinite}. This means that for $x_3>d$, we have the solution: 
\begin{equation}
\psi^+({\bf x}) = \phi_{\rm t}({\bf x})+\phi_{\rm d}({\bf x}),
\label{solution-finite}
\end{equation}
where $\phi_{\rm t}({\bf x})$ is the transmitted (forward-diffracted) wave, with wave vector ${\bf k}_0$, and $\phi_{\rm d}({\bf x})$ the diffracted wave, with wave vector ${\bf k}'_0$, given by: 
\begin{eqnarray}
\phi_{\rm t}({\bf x}) &\approx& \left[ \sum_{\sigma =1,2} \tilde{u}_\sigma\, e^{i(k_{\sigma,3}- k_{0,3})d} \right] e^{i{\bf k}_0 \cdot{\bf x}},\\
\phi_{\rm r}({\bf x}) &\approx& \left[ \sum_{\sigma =1,2} \tilde{u}_\sigma X_\sigma e^{i(k_{\sigma,3}- k'_{0,3})d} \right] e^{i{\bf k}'_0 \cdot{\bf x}}.
\label{solution-finite2}
\end{eqnarray}
The amplitudes $\tilde{u}_1$ and $\tilde{u}_2$ being given by (\ref{solution-u3}), we can write: 
\begin{equation}
\phi_{\rm t}({\bf x}) \approx A_{\rm t}\, e^{i{\bf k}_0\cdot{\bf x}},\quad \phi_{\rm r}({\bf x}) \approx A_{\rm r}\, e^{i{\bf k}'_0\cdot{\bf x}},
\label{solution-finite3}
\end{equation}
with the coefficient $A_{\rm t}$ given by: 
\begin{equation}
A_{\rm t}= {1\over 2}\left[\left(1 - {y\over \sqrt{1+y^2}}\right)e^{{i\over 2}(k_{1,3}- k_{2,3})d}
+\left(1 + {y\over \sqrt{1+y^2}}\right) e^{-{i\over 2}(k_{1,3}- k_{2,3})d}\right] e^{i({1\over 2}(k_{1,3}+ k_{2,3})-k_{0,3})d}.
\label{At}
\end{equation}
Defining the characteristic length 
\begin{equation}
\Delta \equiv {2\pi \sqrt{k_0^2-{1\over 4}g_1^2}\over |\tilde{U}({\bf g})|},
\label{delta}
\end{equation}
from (\ref{detzero-solutions2}) we obtain: 
\begin{eqnarray}
k_{1,3}- k_{2,3} &=& - {2\pi\over \Delta} \sqrt{1+y^2},\label{k-k}\\
k_{1,3}+ k_{2,3} &=&2\sqrt{k_0^2-{1\over 4}g_1^2}-{\tilde{U}({\bf 0})\over \sqrt{k_0^2-{1\over 4}g_1^2}}.
\label{kk}
\end{eqnarray}
If ${\bf g}$ is a vector in the reciprocal lattice, it follows from (\ref{pseudopotential}) that $\tilde{U}({\bf 0})=|\tilde{U}({\bf g})|$, so that (\ref{kk}) becomes: 
\begin{equation}
k_{1,3}+ k_{2,3}=2\sqrt{k_0^2-{1\over 4}g_1^2}-{2\pi\over \Delta}.
\label{k13}
\end{equation}
Also, (\ref{ko1}) gives: 
\begin{equation}
k_{0,3}=\sqrt{k_0^2-{1\over 4}g_1^2}+y{\pi\over \Delta}.
\label{k03-bis}
\end{equation}
Inserting (\ref{k03-bis}), (\ref{k13}) and (\ref{k-k}) into (\ref{At}), we finally obtain:
\begin{equation}
A_t = \left[\cos\left({\pi d\over \Delta}\sqrt{1+y^2} \right)+ {iy\over \sqrt{1+y^2}}\sin \left({\pi d\over \Delta}\sqrt{1+y^2}\right) \right]e^{-{i\pi d\over \Delta}(1+y)}.
\label{At}
\end{equation}
By a similar calculation, we also find for the refracted amplitude: 
\begin{equation}
A_r=-{i\over \sqrt{1+y^2}}\sin \left({\pi d\over \Delta}\sqrt{1+y^2}\right) e^{-{i\pi d\over \Delta}(1+y)},
\label{Ar}
\end{equation}
and it is easy to check that: $|A_t|^2 + |A_r|^2=1$.

\section{Neutron passing through a homogeneous magnetic field}
\label{magnetic}

Having solved the problem of the passage of a neutron through a crystal plate, in the approximation of a weak neutron-nuclei interaction, we consider in this section the passage of a neutron through a homogeneous magnetic field. Also in this case, we  limit our calculation to the situation of a weak field, and will discuss later the situation of a strong magnetic field.

For this, it will be sufficient to consider a one-dimensional problem. We assume that the magnetic field is localized in the interval $[-a,a]$, along the $x$-axis (which corresponds to the direction of propagation of the neutron), and that the magnetic field points in the $z$-direction. Then, the Hamiltonian is $H=-{\hbar^2\over 2m}\partial_x^2 -\mu B(x)S_z$, where $\mu$ is the magnetic moment of the neutron (which is negative), $S_z={\hbar\over 2}\sigma_z$ is the $z$-component of the spin operator, and $B(x) = B \chi_{[-a,a]}(x)$ describes the magnetic field, with $\chi_{[-a,a]}(x)$ the characteristic function of the interval $[-a,a]$. 

So, denoting $\psi_\pm(x)$ the two spin components of the neutron's spin, along the $z$-direction, they are solution of the Schr\"{o}dinger equation:
\begin{equation}
\left[-{\hbar^2\over 2m}\partial_x^2 \pm {\hbar \omega\over 2} \chi_{[-a,a]}(x) \right]\psi_\pm(x) = E\psi_\pm(x),
\label{Schr}
\end{equation}
where we have defined the pulsation $\omega = -\mu B$ and $E={\hbar^2k^2\over 2m}$ is the energy of the incident neutron. So, the ``up'' spin component sees a square potential barrier of height ${\hbar \omega\over 2}$, whereas the ``down'' spin component sees a square potential well of depth ${\hbar \omega\over 2}$.

The solution for a neutron coming from the left requires the boundary conditions: 
\begin{eqnarray}
\psi_\pm(x)&=& e^{ikx} + R_\pm e^{-ikx}, \quad x<-a\nonumber\\
\psi_\pm(x)&=& A_\pm e^{ik_\pm x} + B_\pm e^{-ik_\pm x}, \quad -a<x<a\nonumber\\
\psi_\pm(x)&=& T_\pm e^{ikx}, \quad x>a,
\label{boundarycondition}
\end{eqnarray}
where we have defined $k_\pm = \sqrt{k^2 \mp {m\over\hbar}\omega}$. The continuity of the wave function and of its derivative at the points $x=-a$ and $x=a$ then gives, after some algebra (see any good book of quantum mechanics): 
\begin{eqnarray}
T_\pm &=& {4kk_\pm e^{-2ika}\over (k+k_\pm)^2e^{-2ik_\pm a}-(k-k_\pm)^2e^{2ik_\pm a}},\\
R_\pm &=& {2i(k^2-k_\pm^2)e^{-2ika} \sin(2k_\pm a) \over (k-k_\pm)^2e^{2ika}-(k+k_\pm)^2e^{-2ika}},
\label{TR}
\end{eqnarray}
and of course $|R_\pm|^2+|T_\pm|^2=1$. 

At this point, we assume that the neutron's energy is much bigger than its magnetic energy, i.e., $\hbar\omega \ll E$. Since 
$k^2-k_\pm^2 = \pm{m\over\hbar} \omega$, $k_\pm\approx k(1\mp{\hbar\omega\over 4E})$, and considering that: 
\begin{equation}
{R_\pm\over T_\pm}= -{i(k^2-k_\pm^2)\sin(2k_\pm a) \over 2kk_\pm},
\label{RfrattoT}
\end{equation}
we immediately see that the reflection by the magnetic field is negligeable with respect to transmission, and we have the approximation:
\begin{equation}
T_\pm \approx e^{2i(k_\pm -k)a}\approx e^{\mp i {\omega\over 2} T}.
\label{Tapprox}
\end{equation}

Therefore, if $\xi_+$ and $\xi_-$ are the ``up'' and ``down'' components of the spinor state of the neutron, with respect to the $z$-direction of the magnetic field,  the average of the spin operator $S_\pm = S_x\pm i S_y$ in the transmitted state is:
\begin{equation}
\langle S_\pm \rangle^{\rm tr} = \xi_{\pm}^*\xi_{\mp} T_{\pm}^* T_{\mp} \approx \langle S_\pm \rangle^{\rm in} e^{\pm i2(k_- -k_+)a},
\label{spm}
\end{equation}
where $\langle S_\pm \rangle^{\rm in}=\xi_{\pm}^*\xi_{\mp}$ is the average over the incoming state. This means that by passing through the magnetic field, the spin vector $(S_x,S_y)$ is rotated by an angle:
\begin{equation}
\alpha = 2(k_- -k_+)a\approx \omega T,
\label{alpha}
\end{equation}
where $T={2a\over v}$ is the \emph{sojourn time} of the neutron in the magnetic field region  and $v={\hbar k\over m}$ its speed. In other terms, in the weak magnetic field approximation, the rotation angle of the spin vector is in accordance with classical \emph{Larmor precession}.

\section{Interference effects and the spinor $4\pi$-periodicity}
\label{interference}

In the following we will assume for simplicity that the Bragg condition is perfectly satisfied, i.e., that the incident angle $\theta_0 = \theta_B$, implying that the $y$-parameter is equal to zero. Eqs. (\ref{At}) and (\ref{Ar}) then become: 
\begin{equation}
A_t = e^{-i\tau}\cos\tau, \quad A_r =-ie^{-i\tau}\sin \tau,
\label{AtAr}
\end{equation}
where we have set $\tau\equiv {\pi d\over \Delta}$.
\begin{figure}[!ht]
\centering
\includegraphics[scale =0.22]{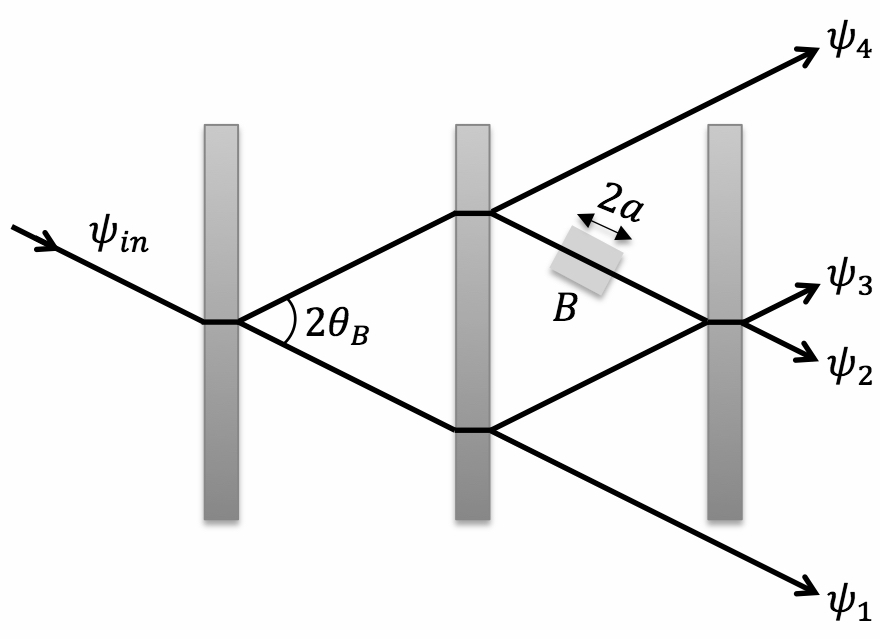}
\caption{A 2-dimensional sketch of the LLL crystal with the incident beam $\psi_{\rm in}$ and the four outgoing beams $\psi_1$, $\psi_2$, $\psi_3$ and $\psi_4$. Only beams $\psi_2$ and $\psi_3$ result from a recombination of the northern and southern paths in the crystal. In the northern path, a local magnetic field of intensity $B$ and spatial extension $2a$ is applied, that by changing the phase of the neutron's wavefunction will give rise to observable interference effects. 
\label{Magnetic}}
\end{figure}
As evidenced in Fig.~\ref{Magnetic}, the incoming wave function $\psi_{\rm in}$ is split into four components $\psi_1$, $\psi_2$, $\psi_3$ and $\psi_4$, each of which arrives in a specific neutron detector (see  also Fig.~\ref{Monolitic}). For the incoming wave we can write: 
\begin{equation}
\psi_{\rm in}({\bf x})= \begin{bmatrix} \xi_+\\ \xi_- \end{bmatrix} e^{i{\bf k}_0\cdot{\bf x}},\quad |\xi_+|^2+|\xi_-|^2=1,
\label{psi-in}
\end{equation}
where $\xi_+$ and $\xi_-$ are the ``up'' and ``down'' components of the spinor state of the neutron, with respect to the direction of the magnetic field (placed in the northern path, inside the LLL crystal; see Fig.~\ref{Magnetic}). When $\psi_{\rm in}$ encounters the first plate, it gets split into two components, a transmitted (forward refracted) and refracted [reflected by the (220) planes of the silicon crystal] one. 

These two components will spatially separate when traveling in the space between the two plates (of course, this separation only occurs when different incoming plane waves are superposed, to form a wave packet), and when they reach the second plate, in distinct locations, they will be either transmitted or deflected, producing in this way four different components. 

As it can be seen in Fig.~\ref{Magnetic}, the $\psi_1$ component experiences two transmissions, one through the first plate and the other through the second plate, before reaching its detector. On the other hand, the $\psi_4$ component experiences a refraction and a transmission, before reaching its detector. Thus, we can write: 
\begin{equation}
\psi_1 = A_t A_t\psi_{\rm in}, \quad \psi_4 = A_t A_r\psi_{\rm in}.
\label{psi-in}
\end{equation}

The two components $\psi_1$ and $\psi_4$ are clearly non-interfering, as they split apart and do not recombine before being detected. This is not the case for the two components $\psi_2$ and $\psi_3$, which result from the overlapping of the two beams refracted by the second plate at the level of the third plate. Both $\psi_2$ and $\psi_3$ receive contributions from the northern and southern paths inside the LLL crystal. The northern path, however, also experiences the dephasing (due to the spin rotation) produced by the static magnetic field. So, for the $\psi_2$ component we can write: 
\begin{eqnarray}
\psi_2 &=& A_r A_r A_t \psi_{\rm in} + A_t A_r A_r \left[ \begin{array}{cc}
e^{- i {\omega\over 2} T} & 0 \\
0 & e^{ i {\omega\over 2} T} \end{array} \right] \psi_{\rm in}\nonumber\\
&=& A_r^2A_t \left[ \begin{array}{cc}
1+e^{- i {\omega\over 2} T} & 0 \\
0 & 1+e^{ i {\omega\over 2} T} \end{array} \right] \psi_{\rm in}.
\label{psi-in}
\end{eqnarray}
Similarly, for the $\psi_3$ component we have:
\begin{eqnarray}
\psi_3 &=& A_t A_r A_t \psi_{\rm in} + A_r A_r A_r \left[ \begin{array}{cc}
e^{- i {\omega\over 2} T} & 0 \\
0 & e^{ i {\omega\over 2} T} \end{array} \right] \psi_{\rm in}\nonumber\\
&=&A_r\left[ \begin{array}{cc}
A_t^2+A_r^2 e^{- i {\omega\over 2} T} & 0 \\
0 & A_t^2+A_r^2 e^{ i {\omega\over 2} T} \end{array} \right] \psi_{\rm in}.
\label{psi-in}
\end{eqnarray}

Using (\ref{AtAr}) and (\ref{psi-in}), after a simple calculation we find that the probabilities of detecting the neutron at detector $D_2$ and $D_3$, placed in the direction of the beams associated with the $\psi_2$ and $\psi_3$ outgoing components of the wave function, are given by:
\begin{eqnarray}
\|\psi_2\|^2 &=& {1\over 2}\sin^2\tau\sin^2 2\tau \left(1+\cos {\omega T\over 2} \right),\\
\|\psi_3\|^2 &=& \sin^2\tau\left(\cos^4\tau + \sin^4\tau -{1\over 2} \sin^2 2\tau \cos {\omega T\over 2} \right)\nonumber.
\label{psi-in}
\end{eqnarray}
We observe that, as expected, the sum $\|\psi_2\|^2 +\|\psi_3\|^2=\sin^2\tau$ does not depend on the spin rotation angle $\omega T$, considering that $\|\psi_1\|^2 +\|\psi_4\|^2 =\cos^2\tau$, in accordance with the conservation of probability (as the Si crystal has essentially zero absorption for thermal neutrons). 

\begin{figure}[!ht]
\centering
\includegraphics[scale =0.2]{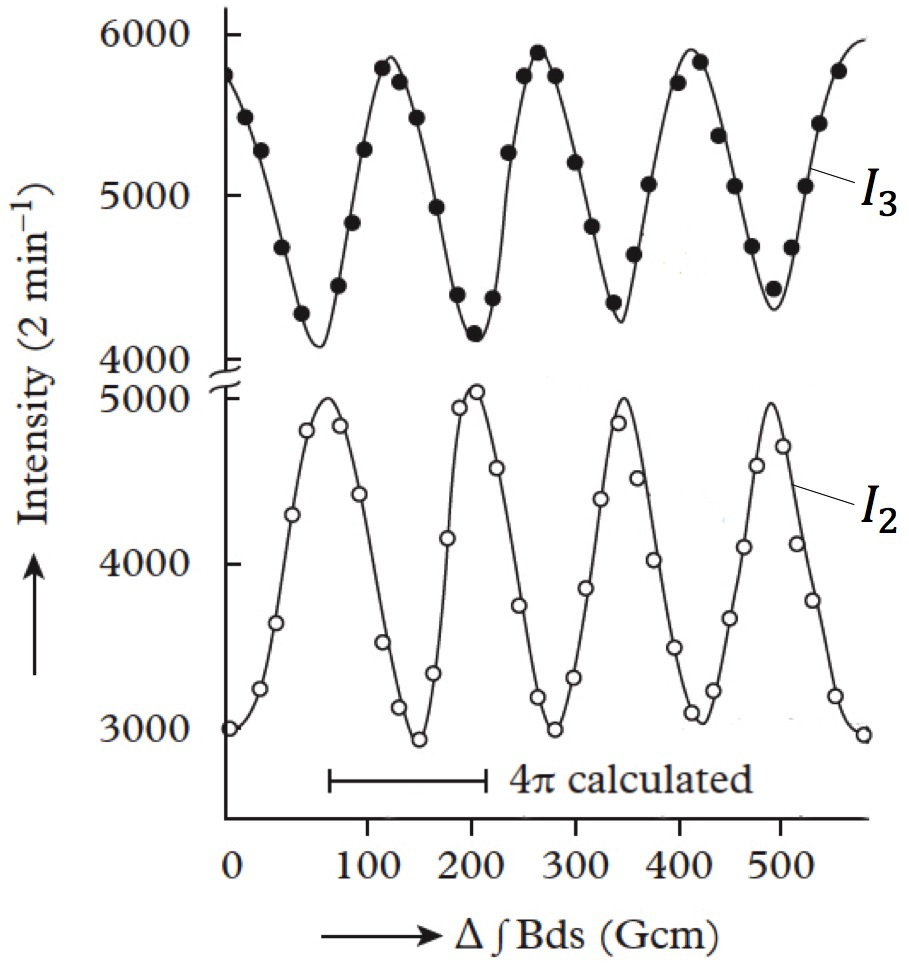}
\caption{The data of the $4\pi$-symmetry experiment performed by Rauch \emph{et al} \cite{Rauchetal1975}, showing the $4\pi$-periodicity of the interfering intensities $I_2$ and $I_3$, when the intensity of the magnetic field is varied. The figure has been adapted from \cite{Rauchetal1975}.
\label{fourpi}}
\end{figure}

So, if we denote $I_0$ the intensity of the incoming neutrons' beam (typically a few thousand counts per minute in Rauch's experiment), then the intensities $I_2$ and $I_3$ at the $D_2$ and $D_3$ detectors will be given by $I_2= I_0\|\psi_2\|^2$ and $I_3= I_0\|\psi_3\|^2$, respectively. Thus, they will exhibit a remarkable $4\pi$-periodicity with respect to the rotation angle $\alpha =\omega T$, as it has been possible to confirm in the celebrated experiments by Rauch \emph{et al} \cite{Rauchetal1975} and Werner \emph{et al} \cite{Werneretal1975}, performed in 1975, as illustrated in Fig.~\ref{fourpi}.

\section{Larmor precession or Stern-Gerlach effect?}
\label{larmor}

According to classical mechanics, any object with a magnetic moment, interacting with an external static magnetic field, will experience a precession of its angular momentum about the magnetic field direction, a phenomenon called \emph{Larmor precession}. 
It can then be shown that the precession angular frequency $\omega$, called the \emph{Larmor frequency}, is given by $\omega = -\mu B$. 

According to this classical picture, a magnetic moment interacting with the magnetic field for a time interval $T$ will experience a total rotation of an angle $\alpha = \omega T$. Clearly, if the magnetic moment is associated with a classical point particle moving in space, with velocity $v$, the interaction time $T$ will be given by the time spent by the particle inside the field region, i.e., $T={2a\over v}$. This is precisely the rotation angle for the neutron's spin that we have calculated in Sec.~\ref{magnetic}, by solving the Schr\"{o}dinger equation in the low-field limit. So, can we conclude that the phase shifts of the ``up'' and ``down'' spin components result from a classical-like Larmor precession?

There has been a little controversy in the literature in relation to this question \cite{Mezei1979,Bernstein1980,Mezei1986,Barut1987,Mezei1988,Barut1990,Gough1992,Martin1992,Martin1994}. Indeed, there is an important difference between a magnetic moment associated with a classical point particle and a magnetic moment associated with a quantum ``particle,'' as the latter, when subjected to a magnetic field gradient, will generally experience a Stern-Gerlach effect. In the standard Stern-Gerlach experiment, only the gradient of the magnetic field perpendicular to the velocity of the quantum entity is taken into consideration, but of course there is also a gradient parallel to it, producing an additional longitudinal (instead of transverse) effect. 

In the experimental set up described in Fig.~\ref{Magnetic}, the only Stern-Gerlach effect to be considered is the longitudinal one; it will cause the wave packets associated with the ``up''/``down'' spin components to be decelerated/accelerated when entering the magnetic field region, so that they will travel inside of it at different speeds $v_\pm ={\hbar k_\pm\over m}$, so producing their spatial separation, a phenomenon which has been clearly observed in specific experiments \cite{Alefeld1981,Weinfurter1981}. This means that the answer to the above question is negative: the effect of passing through the magnetic field region is not that of inducing a Larmor precession of the neutron's spin, but a space distancing of the wave-packets associated with its two components. 

Of course, for as long as one just considers a weak magnetic field, as we did in the calculation of Sec.~\ref{magnetic}, it is impossible to distinguish a Larmor precession from a longitudinal Stern-Gerlach effect, as in the weak field limit the phase change produced by the latter is identical to that produced by the former. So, to truly understand what is the physical mechanism underlying the phase change one needs to analyze the situation beyond the weak field limit. 

In this case, however, if the switching on and off of the field in space is abrupt, part of the wave packet of the neutron will also be reflected. To avoid this additional complication, one simply has to consider a magnetic field $B(x) = B w(x)$, where the function $w(x)$, $0\leq w(x)\leq 1$, describing the spatial localization of the field, is taken to be sufficiently smooth (differentiable), as in this case it can be shown that the reflection effects can be neglected, even if the field is not weak (provided of course that $E>B$). In particular, if the switching on and off of the field in space is assumed to occur on a much larger scale than the wavelength of the neutron ($k \gg dw(x)/dx$), then, by using semiclassical methods, it is possible to show that we have the approximation \cite{Martin1992,Martin1994}:
\begin{equation}
T_\pm \approx e^{i\int_{-\infty}^\infty dx [k_\pm(x) -k]},
\label{Tapprox2}
\end{equation}
where $\hbar k_\pm(x)=\sqrt{2m(E\pm Bw(x)}$.

This means that, when passing through the magnetic field region, the neutron will be rotated by an angle
\begin{equation}
\alpha =\int_{-\infty}^\infty dx\, \left[k_-(x) -k_+(x)\right].
\label{alpha2}
\end{equation}
To see if $\alpha$ can still be interpreted as a precession angle, one should be able to write it as $\alpha \approx \omega T(B)$, with $T(B)$ the sojourn time of the neutron inside the magnetic field region. This is easy to check, as there is a well defined notion of \emph{sojourn time} (also called dwell time) in quantum mechanics \cite{Martin1981,Hauge1989,Sassoli2012}. We shall not perform here the calculation of the sojourn time, as this would go beyond the scope of the present article, and refer the interested reader to \cite{Martin1981,Martin1992,Martin1994}. Let us however comment the result. 

For this, we consider a more explicit form for the cut-off function $w(x)$. We assume that $w(x)=1$, if $x\in [-a,a]$, and $w(x)=g({|x|-a\over l})$ otherwise, where the parameter $l>0$ is a measure of the size of the field gradient [since $dw(x)/dx=O(l^{-1})$], and $g(x)$, $x\geq 0$, is a smooth function with compact support, such that $0\leq g(x)\leq 1$, and $g(0)=1$. Then, it is possible to demonstrate that the neutron's sojourn time is approximately:
\begin{equation}
T(B) \approx {2a\over v_+}|\xi_+|^2 + {2a\over v_-}|\xi_-|^2,
\label{T(B)}
\end{equation}
where $v_\pm = {\hbar k_\pm\over m}$ and $k_\pm = \sqrt{2m(E\pm B)}$. 

This result is not surprising: it simply tells us that the time spent on average by the neutron in the field region is given by the weighted sum of the two contributions of the spin components. More precisely, the first contribution is the time spent in the field region by a particle of speed $v_+$, times the spin ``up'' probability $|\xi_+|^2$, and the second contribution is the time spent in the field region by a particle of speed $v_-$, times the spin ``down'' probability $|\xi_-|^2$. This is clearly in accordance with the fact that the field gradient produces and acceleration ($v\to v_+$) of the ``up'' component and a deceleration ($v\to v_-$) of the ``down'' components, which will then propagate ad different speeds inside the magnetic field. 

The important point for our discussion is that $T(B)$ depends on the initial spin state, via the probabilities $|\xi_+|^2$ and $|\xi_-|^2$. On the other hand, the rotation angle (\ref{alpha2}) does not depend on the initial spin state. Thus, it is impossible to interpret (\ref{alpha2}) as being the result of a Larmor precession, beyond the weak field regime; and of course, the reason why when $B\to 0$ this interpretation is possible, is that in this limit $v_\pm\to v$, and since $|\xi_+|^2 + |\xi_-|^2 =1$, $T(0)\approx{2a\over v}$ becomes independent of the spin state.

\section{Is a $2\pi$-rotation observable?}
\label{rotation}

In the previous section, we have shown that the rotation angle of a neutron passing through a static magnetic is the result of a longitudinal Stern-Gerlach effect, and not of a Larmor precession. However, for what concerns the problem of the observability of the $4\pi$-symmetry, it is rather irrelevant if the magnetic field acts primarily on the translational degrees of freedom of the neutron, or on the ``internal'' spin-rotational ones: what is important is that such action happens in a way that ultimately produces a rotation of the incoming spin, which is what happens in the weak field limit (when the rotation angle tends to zero), but also for a finite magnetic field strength, when the spatial switching on and off occurs on a much larger scale than the neutron's wavelength, as also in this case there are no reflections at the field boundaries and one obtains (at least in principle) a finite rotation angle with no polarization effects. 

In this section, we briefly discuss the issue of the observability of the $e^{-i\pi}=-1$ phase change of the state of the neutron's spin when rotated through $2\pi$. The question is: If we $2\pi$-rotate a spin-${1\over 2}$ entity, can this produce an observable effect? Apparently, this is precisely what the experiments by Rauch \emph{et al} \cite{Rauchetal1975} and Werner \emph{et al} \cite{Werneretal1975} have done, as indeed the measured intensities of the two interfering beams exhibit a remarkable $4\pi$-periodicity (see Fig.~\ref{fourpi}). However, one should not confuse the rotation of an entity with the rotation of part of that entity, as in the latter case it is perfectly reasonable to expect that a $2\pi$-rotation will produce an observable effect. 

\begin{figure}[!ht]
\centering
\includegraphics[scale =0.16]{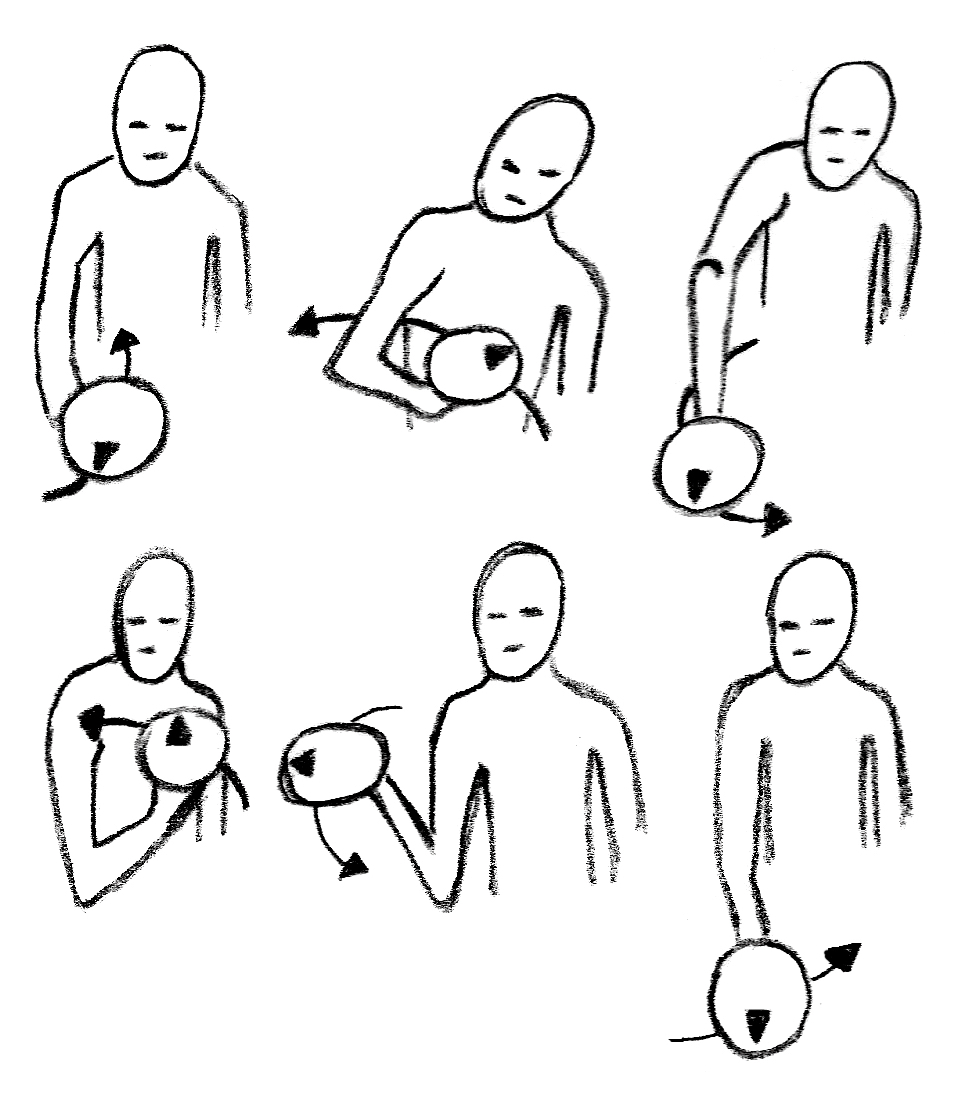}
\caption{Two full rotations of the plate are needed to bring the arm and the plate in the same initial condition. 
\label{Binasuan}}
\end{figure}

A typical example is the traditional Philippine dance called \emph{Binasuan}, where dancers hold full wine glasses in each palm of the hand, while rotating the latter, over and under their elbows, to keep their palms always facing up (to avoid spilling the liquid). An example of these movements is sketched in Fig.~\ref{Binasuan}, with the glass replaced by a plate. We can observe that when the plate is $2\pi$-rotated around the vertical axis (the first three drawings of Fig.~\ref{Binasuan}), this produces a visible twist in the arm. However, such twist can be dissolved, and the arm can regain its initial non-twisted condition, if the plate is further rotated by $2\pi$ (the last three drawings of Fig.~\ref{Binasuan}).

Of course, the plate on the palm of the hand is just an extension of the latter, allowing to better visualize its rotational movement and the fact that it always remains palm up, during the entire $4\pi$-rotation, i.e., the (vertical) rotation axis remains fixed (in the same way as the magnetic field direction remains fixed in the experiment with neutrons). Now, the reason why the $2\pi$ rotation of the hand doesn't bring the body-entity back to the same condition is that the hand is not an isolated entity, but part of a bigger entity -- the entire body of the person -- to which it is connected via the arm; and the geometry of this connection is such that only when performing two full rotations the body-entity can recover its exact initial condition. 

In the experiment with neutrons the situation is at the same time similar and different. It is similar because it is not the entire wavefunction that is rotated, but only that part which goes through the magnetic field region, whereas its other parts are not at all affected by the latter (because of the great distance between the beams inside the interferometer). 

So, in the same way that the movement depicted in Fig.~\ref{Binasuan} has nothing to do with a rotation of the entire body-entity, the passage of a neutron through the LLL interferometer does not produce a rotation of the latter, but just a rotation of part of it; and in the same way that the $2\pi$-rotation of the hand of a Binasuan dancer cannot bring its entire body in the same configuration, the same is also to be expected when just ``a hand'' of the neutron is being $2\pi$-rotated. The situation with neutrons is however also different, because neutrons cannot be treated a entities `locally existing in space', as we will better discuss in the next section. 

It is important to observe that the speed of the incoming neutrons, for instance in Rauch \emph{et al} experiment \cite{Rauchetal1975}, was of about $2.2\cdot 10^3$~m/s (approx. 5000 miles per hour), and that the distance between them was on average of $300$ m. Therefore, for all practical purposes, there was just a single neutron at a time passing through the interferometer. This means that what was measured is not the phase change of neutrons passing through the magnetic field region relative to the phase of other neutrons not passing through it, by means of a coherent superposition. What was observed is  a genuine self-interference phenomenon, where part of an individual neutron's wave function is put out of phase with respect to another part of it, by means of a local interaction. And since such local interaction depends on the neutron's spin via the term $-\mu B(x) S_z$ in the Hamiltonian, and that for a neutron the spin eigenvalues are $\pm{\hbar\over 2}$, the presence of these one-half factors will necessarily introduce a $4\pi$-symmetry in the phase shifts produced by the magnetic interaction, and in the associated interference phenomena. 

So, for what concerns the question of the observability of a $2\pi$-rotation of a neutron, and more particularly of its spin, if we take seriously the idea that a rotation is such only if the entire entity in question is rotated, then the answer is clearly negative (a conclusion that also follows from the very formalism of quantum mechanics; see for instance \cite{Hegerfeldt1968}). Indeed, when in an experimental situation we measure the properties of a given single neutron, this means that the experimental context is such that that neutron will be temporary separated (for all practical purposes) from the other entities, and can be considered as an isolated system. This means that we can attach a Hilbert space of states to it, and that its state, in that experimental context, will be given by a one-dimensional projection operator $\rho=|\psi\rangle\langle\psi |$; and of course, a global phase factor, associated for instance with a $2\pi$-rotation of the \emph{whole} neutron entity, will not change the state, and therefore will not be directly observable.

\section{Non-locality \& non-spatiality}
\label{non-spatiality}

\begin{figure}[!ht]
\centering
\includegraphics[scale =0.3]{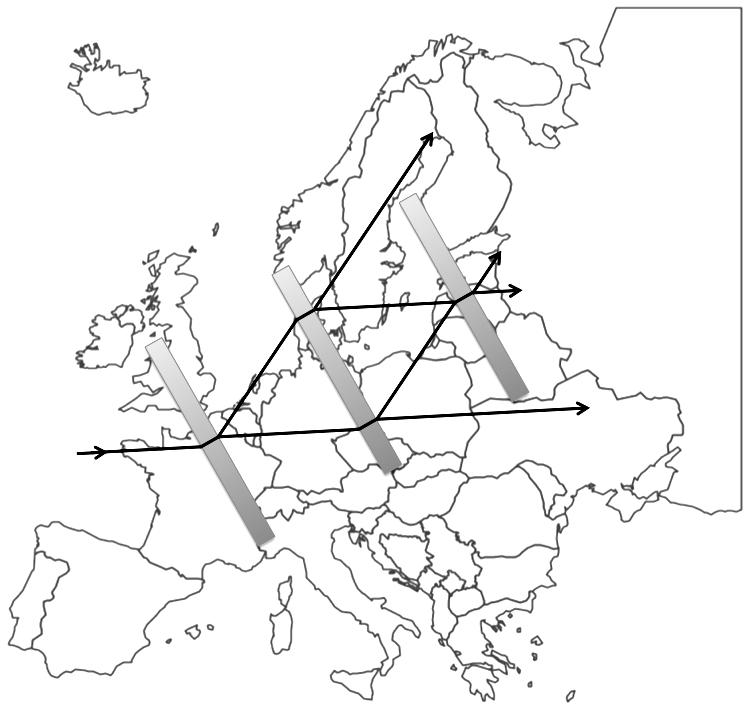}
\caption{The LLL crystal scaled up 25 millions times and projected on the European map.
\label{Poland}}
\end{figure}

Having clarified how the $4\pi$-periodicity of the intensities of the outgoing beams should be interpreted, we want now to analyze the experiment from the viewpoint of the notion of non-locality and its possible interpretation. To do so, it is instructive, following \cite{Aerts1999}, to first rescale the experiment, to appreciate how widely separated are the two coherent beams in the interferometer, which the experimenters is able to manipulate individually. More precisely, scaling up the LLL crystal 25 million times, then projecting it on a European map, we see that the neutron enters the first plate in France, close to Paris, and that once it has overpassed the second plate, the northern beam crosses Sweden, whereas the southern one passes over Poland, before being both recombined in Latvia (see Fig.~\ref{Poland}). 

In Rauch's experiments neutrons have a typical `longitudinal coherence length' of one millionth of a centimeter. Scaling up this length also 25 million times, it means that the region within which this imaginary giant neutron can be acted upon, when it travels along its different possible paths, is a small cube of only 25 centimeters. This of course indicates that the neutron is not an extended spatial object (like a wave), but really something like a small projectile that would have been fired towards the crystal from the Atlantic Ocean, and would then travel either to the north, to reach Sweden, or to the south, to reach Poland, which are hundreds of kilometers away from one another (for simplicity, we do not consider in the discussion the two non-interfering paths). 

However, this `projectile view' cannot be correct, and this precisely because of the observed $4\pi$-periodicity of the interference effects. Indeed, if neutrons would really be like localized projectiles, then either they would not encounter the magnetic field, when passing through Poland, or they would pass through it, when reaching Sweden, causing their spin to be globally rotated by it. But in none of these two cases we would observe interference effects, as the detectors $D_2$ and $D_3$ (see Fig.~\ref{Monolitic}) would then click on average exactly the same number of times, independently of the presence of the local magnetic field in the northern path. 

To make this argument even more stringent, one can consider that the intensity of the magnetic field has been chosen so as to exactly rotate the neutron's spin of $2\pi$. Then, as we discussed in the previous section, if the whole spin entity is $2\pi$-rotated (assuming that it is fully localized in the region of the magnetic field), it will be back to exactly the same state, so that no effects should be observed at the level of the outgoing intensities $I_2$ and $I_3$. However, experiments tell us something completely different. 

Thus, we must conclude that the view of a neutron as a spatial projectile is necessarily false, as certainly it does not travel along the northern or southern paths in \emph{actual} terms. And of course, it also cannot be jointly travelling along both paths, in actual terms, as this would correspond to the situation of having two neutrons at the same time in the interferometer; and as we said already, we cannot either explain the interferences by assuming that the neutron would be an entity spread out in space, with an extension of hundreds of kilometers (on the scale where the interferometer covers half of Europe). This not only because of its very small coherence length, but also because if we would place detectors in the space between the northern and southern paths, they would never click, and if we would place detectors in correspondence of the two paths, either they would detect a complete neutron, or no neutron at all (the situation where the two detectors would jointly detect a `portion of a neutron' being never observed).

So, neutron interferometry experiments force us to go beyond the na{\"\i}f wave-particle duality, since both particles and waves remain local entities, whereas neutrons are genuinely \emph{non-local} (or `de-localized'), i.e., they are entities that can be prepared in a state such that they can be influenced by macroscopically separated regions of space by local apparatus (like a magnetic field) acting only in one (or several) of these separated regions at one time, but without manifesting in between these apparatuses and still remaining whole local entities when they are detected \cite{AertsReigner1991}.

What can we say, then, about the nature of neutrons? How can an entity be influenced from space without actually being localizable in space. Is this even logically possible? Prof. Constantin Piron, who I mentioned in the Introduction, in his celebrated course of quantum mechanics, in Geneva, liked to use the following metaphor \cite{Piron1990}: Take a dollar bill. When it is intact, we can say that a dollar is located somewhere in space, like a classical entity. But what happens if the bill is torn in two parts and the obtained two pieces are spatially separated and, say, put each one in a different box? In this situation, we cannot say that the dollar is still located somewhere in space: it has disappeared from space, but also not completely. We can still say that, in a sense, the dollar is present in the two boxes, but that at the same time it is clearly contained in none of them, quite similarly to a neutron that would be simultaneously present in all its possible paths, although certainly not in actual terms. The two boxes, when taken together, contain one dollar, but when considered separately they don't. All we can say is that they both contain a \emph{potential} dollar, which can become actual only when the two pieces of bill are recombined. 

The dollar bill is of course only a metaphor, but an interesting one, as it conveys two very important ideas. The first one is that a microscopic quantum entity would be non-local because it would be \emph{non-spatial}. By non-spatial we don't mean here an entity that would have totally disappeared from space, here understood as a theater describing the relations between the macroscopic entities of our human experience. Indeed, if this would be the case, then it would be impossible to understand why a quantum entity can be influenced by classical entities, like measurement apparatuses, present in space. 

So, a microscopic quantum entity, like a neutron, cannot be considered to be present in space, but nevertheless still maintains a relation with space, by always remaining available to be detected inside of it, when certain conditions are met; and the locations in space where the `degree of availability' (i.e., the probability) of the neutron to be detected is non-negligible (along the four different possible paths in Rauch \emph{et al} experiment) can be considered as the windows through which we can act on its non-spatial state and have access to its ``pre-spatial'' reality \cite{AertsReigner1991}.

This view, of a neutron being typically a non-spatial entity, with non-spatiality being the truly mysterious aspects in Rauch \emph{et al} experiments, conveys, or is related to, another fundamental view, called the \emph{creation-discovery} view. Quoting from \cite{Aerts1999}: ``Within the creation-discovery view it is taken for granted that during an act of measurement there always exist two aspects, a discovery of a part of reality that was present independently of the act of measurement, and a creation that adds new elements of reality to the process of measurement and to the entity under investigation.'' In neutron interferometry experiments the main `creation aspect' is of course the `creation of a place in space' for the neutrons, when they are detected, as before being detected they remain just potentially (and not actually) present in space. 

Coming back to Piron's one-dollar example, the reason it works well with our intuition is that a one-dollar bill is not only an \emph{object}, but also a \emph{concept}. When we say that the bill is in both boxes, and that at the same time is in none of them, this still makes sense because two different one-dollar notions exist simultaneously in our minds. There is the concrete `one-dollar bill', which is an objectual entity, and there is the `one-dollar' as a conceptual entity, which can be instantiated (i.e., objectified, concretized) in different ways, like for instance by means of a full banknote. Therefore, when we say that the dollar is not in any of the two boxes, we are more precisely referring to the objectual one-dollar bill. On the other hand, when we say that it is in both boxes simultaneously, we are referring to the dollar as an abstract entity, that can be instantiated in different ways, in the different contexts. 

In the same way, it has been suggested by Aerts that this interplay between the abstract and concrete levels in our human cultural reality could also be at play, \emph{mutatis mutandis}, in the physical world, in the sense that although an entity like a neutron has clearly nothing to do with a human concept, it would nevertheless behave in a conceptual way, that is, in a way that is similar to how human concepts combine together, to produce emergent meanings, and interact with memory structures that are sensitive to their meanings. 

In that sense, when using their measuring apparatuses, physicists would not exactly `watch' nature, but more precisely would 'speak' with nature, and by doing so they would always create (actualize) new correlations, properties and meanings. Of course, explaining how Aerts' \emph{conceptuality interpretation} can clarify some of the most puzzling aspects of quantum theory, like non-locality/non-spatiality, entanglement, interference and superposition, identity and individuality, would go beyond the scope of the present article, and for this we refer the interested reader to \cite{Aerts2009,Aerts2010}.

\section{Conclusion}
\label{conclusion}

To conclude, let us briefly summarize what we have presented in this article. Firstly, we have provided a full treatment of the passage of a neutron through a LLL interferometer. We have performed the calculation by only considering expressions at the first order in the interaction of the neutron with the nuclei, so that it was possible to neglect the backward reflection effects by the crystal's plates; and of course, we have also neglected in our calculation the magnetic interaction of the neutron with the crystal's atoms (the coupling between the momentum and spin variables would have produced an additional small splitting of the neutron propagation directions in the crystal; see \cite{ZeilingerShull979}).

Secondly, we have considered the problem of the passage of a neutron through a static magnetic field, and we have analyzed the origin of the spin rotation, which apparently is not due to a Larmor precession, but to a longitudinal Stern-Gerlach effect. This means that when a neutron goes in and out from a region where a magnetic field is applied, classical properties would not show up, primarily, in the spin evolution, but in the translational motion, with the ``down'' wave-packet component being delayed with respect to the ``up'' one.

Combining these two analysis, we have then provided explicit expressions for the outgoing intensities measured by the detectors, showing that those associated with the two interfering beams are $4\pi$-periodic with respect to the rotation angle of the neutrons' spin components passing through the magnetic field, in accordance with the experimental data.

Finally, we have emphasized that experiments with neutrons do not demonstrate the observability of a $2\pi$-rotation, because to really speak of a rotation, instead of a torsion, the entire spinorial entity must be rotated. This has led us to discuss the important notion of non-locality, which is clearly and forcefully demonstrated by the experiments. Our point, in accordance with Aerts, is that when lucidly analyzed these experiments lead to the inescapable conclusion that neutrons, like other `quantons', are genuinely non-spatial entities. (For alternative ways of deducing the non-spatiality of the microscopic, see \cite{
Sassoli2011,Sassoli2012b,Sassoli2013,AertsSassoli2014,asdb2015a,asdb2015b}; see also the interesting discussion in \cite{Kastner2013} about the importance of questioning the idea that something must exist in space to be real). 

It should be said, however, that the idea of a non-spatial (pre-spatial, pre-empirical, etc.) layer of our reality still remains a difficult one to digest for the majority of physicists. That microscopic quantum entities, like neutrons, when not bound to macroscopic objects, would not always reside in space is still perceived as one of those extraordinary claims that, as Carl Sagan used to put it, requires extraordinary evidence. One of the points of our conceptual analysis was to emphasize that, according to the present author, such extraordinary evidence has already been provided by the beautiful neutron interferometry experiments, and of course also by the experiments with entangled pairs that were realized a few years later, by Aspect \emph{et al} \cite{Aspect1}.

 In other terms, we believe that we are today forced to view space not so much as an all embracing setting (a ``container'' for reality), but as an intermediate structure that we humans have in part constructed, relying on our everyday experience with the macroscopic entities, and in which certain ``encounters'' are allowed to occur \cite{Aerts1999,asdb2015b}.

\section*{Acknowledgements}
I would like to thank G\'erard Wanders, professor emeritus of the University  of Lausanne, for having suggested me, when I was a student, to explore the physics of neutron interferometry experiments. Many of the calculations presented in the first part of this article, describing the passage of a neutron through a LLL crystal, are drawn from  some of his handwritten notes. No need to say, possible errors are solely my responsibility.

\end{document}